\documentclass[conference, 10pt]{IEEEtran}

\usepackage[linesnumbered,ruled]{algorithm2e}
\usepackage{amsmath, amssymb, amsthm}
\usepackage{graphicx}
\usepackage[utf8]{inputenc}
\usepackage{flushend}
\usepackage[hyphens]{url}
\usepackage{multirow}
\usepackage{fancyhdr}

\usepackage{setspace} 
\newtheorem{definition}{Definition}
\newtheorem{theorem}{Theorem}
\usepackage{xcolor}

\title{A New Paradigm in Split Manufacturing:\\Lock the FEOL, Unlock at the BEOL}

\author{
	{\large Abhrajit Sengupta $^\dagger$, Mohammed Nabeel $^\ddagger$, Johann Knechtel $^\ddagger$, and Ozgur Sinanoglu $^\ddagger$}\\
  {$^\dagger$New York University, $^\ddagger$New York University Abu Dhabi}\\
  {\{as9397, mtn2, johann, ozgursin\}@nyu.edu}
}
\begin{document}

\maketitle

\renewcommand{\headrulewidth}{0.0pt}
\thispagestyle{fancy}
\pagestyle{fancy}
\cfoot{
	\vspace{-1cm}
\copyright~2019 IEEE.
This is the author's version of the work. It is posted here for your personal use.
	Not for redistribution.\\
	The definitive Version of Record is published in
	Proc. Design, Automation \& Test in Europe (DATE) 2019\\
}

\begin{abstract}
Split manufacturing was introduced as an effective countermeasure against hardware-level threats such as IP piracy, overbuilding, and
insertion of hardware Trojans. Nevertheless, the security promise of split manufacturing has been challenged by various attacks, 
which exploit the well-known working principles of physical design tools to infer the missing BEOL interconnects.
In this work, we advocate a new paradigm to enhance the security for split manufacturing.  Based on Kerckhoff's principle, we protect the
FEOL layout in a formal and secure manner, by embedding keys.
These keys are purposefully implemented and routed through the BEOL in such a
way that 
they become indecipherable to the state-of-the-art FEOL-centric attacks.
{We provide our secure physical design flow to the community.}
We also define the security of split manufacturing formally and provide the associated proofs.
At the same time, our technique is competitive with current schemes in terms of layout overhead, especially for practical, large-scale
designs (ITC'99 benchmarks).

\end{abstract}
\begin{IEEEkeywords}
Split manufacturing, proximity attack, ATPG
\end{IEEEkeywords}

\section{Introduction}

The semiconductor industry has seen a growing reliance on external parties for cost-effective access to advanced
fabrication facilities. However, such sharing of the valuable intellectual property (IP) with potentially
\textit{untrusted} parties has raised several security concerns.
For example, it is estimated that 15\% of all the ``spare and replacement semiconductors'' bought by the Pentagon are counterfeit which can have potential consequences on national security~\cite{pentagon}.

To address such concerns directly at manufacturing time, IARPA advocated a concept called \emph{split manufacturing}~\cite{iarpa,
split1}. The asymmetry between metal layers facilitates splitting the design into two parts;
the
front-end-of-line (FEOL) contains the active device layer and lower metal layers (usually $\leq$ M3), whereas the back-end-of-line
(BEOL) contains the higher metal layers ($\geq$ M4).
The fabrication of the FEOL requires access to an advanced facility and, thus, this part
is outsourced to a high-end but potentially \textit{untrusted} foundry. The BEOL is subsequently grown on top of the FEOL at a
low-end but \textit{trusted} foundry.
Several studies~\cite{split1, split2, sp1} have successfully demonstrated split manufacturing for complex designs, such as for an asynchronous
FPGA.

Recently, the promise of split manufacturing has been called into question by a class of attacks known as \emph{proximity attacks}. Even though
the design can be split into two parts during manufacturing,
the physical design (PD) tools still have to work on the design holistically to be able to apply various optimization techniques.
This fact, in conjunction with the deterministic nature of commercial PD tools, leaves hints in the FEOL layout which can
be exploited to infer the missing BEOL connections.
In particular, to-be-connected cells are placed nearby in the FEOL, 
mainly to minimize delay.
This hint of physical proximity was first explored by Rajendran
\textit{et al.}~\cite{jv-attack13}.
Several other attacks with consideration of further hints have followed, e.g., \cite{wang18_SM, magana16, zhang18}.

Consequently, several studies sought to defend against proximity attacks, e.g.,~\cite{jv-attack13,jaggu14,sengupta_iccad17,magana16,patnaik_aspdac18,patnaik_dac18}.
Nevertheless, almost all prior protection schemes
rely on some heuristics, which can be broadly understood as perturbing
the placement and/or routing of the FEOL. Such schemes have two general pitfalls: (1)~without any formal security guarantees, the
resilience of the schemes can only be evaluated empirically, 
and 
(2)~perturbing the
FEOL layout ``on top'' of regular design optimization tends to induce large overhead.  
Therefore, there is a clear need for formally secure, yet affordable schemes to advance split manufacturing.

To the best of our knowledge, this work represents the first formally secure scheme for split manufacturing concerning the
classical threat model.\footnote{Imeson \emph{et
al.}~\cite{imeson13} were presumably the first to offer a formal notion for split manufacturing in 2013. However, their work considers a
different threat model: the attacker already holds the full netlist and aims for targeted Trojan insertion. In our work,
we consider the classical threat model, where the objective for the attacker is to infer the full netlist from the FEOL.}
We propose and follow a new paradigm for
split manufacturing---lock the FEOL, unlock at the BEOL---where the resilience of the split layout is formally underpinned with a secret key.
Toward this end, we follow Kerckhoff's principle, i.e., our approach remains secure even if all the implementation details are available
\emph{except} the key.

The contributions and structure of this paper are summarized as follows:
\begin{itemize}

\item A new paradigm for split manufacturing is proposed: lock the FEOL, unlock at the BEOL.
The essence of it is to adapt the notion of logic locking to secure the FEOL and to route the nets holding the key through the BEOL.
\item We formally define the security promise of split manufacturing and establish our paradigm with related proofs.

\item We show that the key can be purposefully implemented without providing any hints to an FEOL-centric attacker, hence
hindering any kind of proximity attacks.

\item A physical design
framework is developed for end-to-end security and layout analysis.
The framework serves to (1)~lock the FEOL by embedding a key into it,
(2)~implement the key with TIE cells and route the related nets
through the BEOL, and (3)~control the layout cost.

\item Extensive experiments on ITC'99 and ISCAS benchmarks are carried out to illustrate the efficacy of our scheme in terms of security and
overhead. For an empirical security analysis, we leverage the state-of-the-art proximity attack in~\cite{wang18_SM}.
Besides, we further assume an ``ideal proximity attack,'' providing the most conservative analysis setup, which still cannot break our scheme.
\end{itemize}

\section{Concept and Formal Analysis}\label{sec:design}
Our work is inspired by the notion of logic locking~\cite{epic, sengupta_vts18},
    but there are important differences as detailed next.

\subsection{Threat model}
\begin{figure}[tb!]
	\centering
	\includegraphics[width=.95\columnwidth] {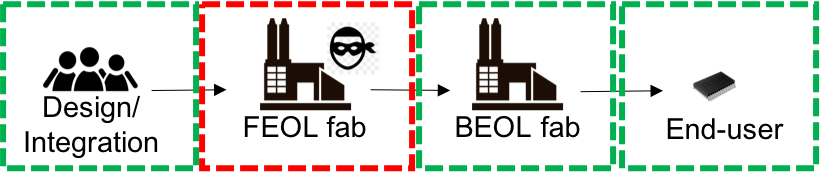}  
	\caption{The threat model for split manufacturing. Red means the entity is untrusted (FEOL fab), whereas green means trusted
		(all other).
}
	\label{fig:threat_model}
\end{figure}

Our threat model (Fig.~\ref{fig:threat_model}), i.e., the classical one for split manufacturing, differs from that for logic locking.
Locking protects against \textit{untrusted end-users} and \textit{untrusted foundries},
whereas split manufacturing can only protect against untrusted FEOL fabs.
This difference implies two consequences:
(1)~For locking, the untrusted end-user forces the designer to store the key in a \textit{tamper-proof memory}.
Such memories remain an active area of research~\cite{tuyls06} and,
hence, their practicality is currently limited.
Additionally, the need for such a memory requires dedicated circuitry such as read/write logic,
which can adversely affect
the layout cost.
(2)~For ours, since the chip fabrication is not yet complete and,
in any case, since the end-user is trusted, there is no physical
oracle available to the attacker. Nevertheless, the attacker has access to the FEOL and full knowledge of the PD tools, the technology library, etc.

\subsection{Physical embedding of the key}

\noindent\textbf{Locking the FEOL.} We secure the layout by inserting additional gates into the design, so-called
\emph{key-gates}.
We lock the entire FEOL by inserting a sufficient number of key-gates that are driven by a secret key.
Toward that end, any locking technique can be applied, including random insertion of key-gates~\cite{epic}.
The conceptional difference to locking, however, is that for us the key is implemented through connections only in the BEOL (as
		opposed to a tamper-proof memory), particularly in a way that is indecipherable to the FEOL-centric attacker applying
proximity attacks.

\begin{figure*}[tb!]
	\centering
	\includegraphics[width=.95\textwidth, trim = {1.5cm 6.8cm 1.3cm 7.8cm}, clip = true] {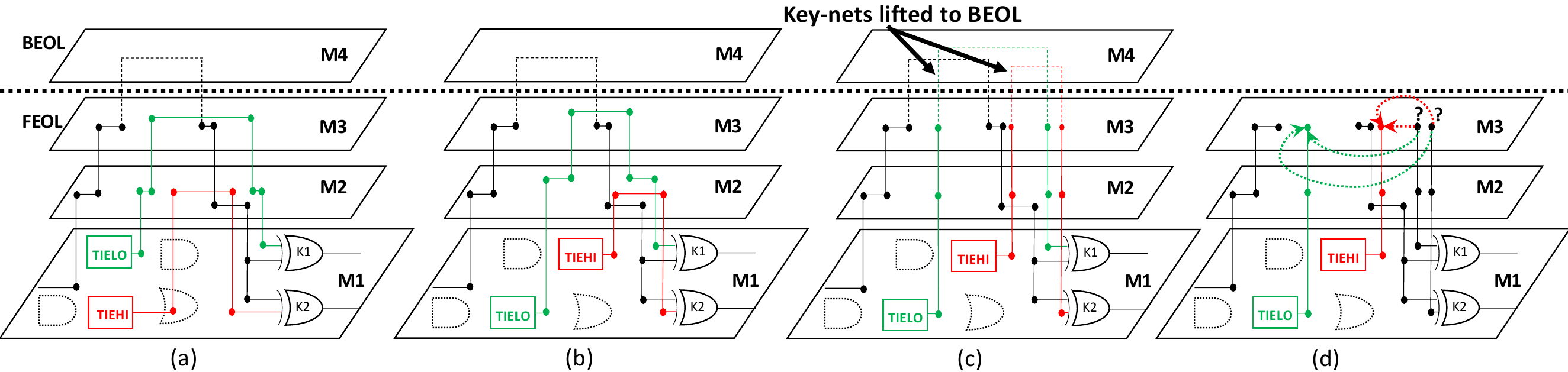}
	\caption{Physical design of the key.
		(a)~Locked layout, with key-nets driven by TIE cells, but with na\"ive physical design.
		(b)~Locked layout of (a), with randomized placement for TIE cells.
		(c)~Locked layout of (b), with key-nets lifted to the BEOL (above M3 here).
		Note that lifting makes use of stacked vias to reduce FEOL routing to the bare minimum.
		(d)~Locked layout of (c) after splitting. {As there are no placement- and routing-related hints remaining for the broken key-nets concerning their assignment to TIE cells, the key is indecipherable for an FEOL-centric attacker.}}
	\label{fig:PD_key}
\end{figure*}

In this work, key-gates are connected with \emph{TIE} cells. These cells provide constant logic 1 and 0, as \emph{TIEHI} and \emph{TIELO} cells, respectively.
A na\"ive physical design of TIE cells and key-gates might
reveal some hints such as proximity-related information
(Fig.~\ref{fig:PD_key}(a)).
Thus, we advocate the following two strategies to ensure security.

\noindent\textbf{Randomized placement of TIE cells.}
To defeat any proximity attack, it is critical that the placement of TIE cells does not reveal any 
connectivity hints. Thus, we propose to
randomize the placements of TIE cells (Fig.~\ref{fig:PD_key}(b)).
We can reasonably expect that doing so has only little impact on layout cost.  TIE cells are very small in comparison to regular cells and,
being no actual drivers, moving TIE cells randomly away from their sinks (the key-gates) does not induce larger loads.

\noindent\textbf{Lifting of key-nets.}
We denote nets connected to TIE cells, representing the key-bits, as {\em key-nets}.
It is easy to see that in case a key-net is fully routed in the FEOL,
the related key-bit can be readily discerned by the attacker (Fig.~\ref{fig:PD_key}(a, b)).
Thus, we advocate a systematic and careful lifting of all key-nets to the BEOL (Fig.~\ref{fig:PD_key}(c)).
Once the design is split, the lifted key-nets form broken connections for an FEOL-centric attacker (Fig.~\ref{fig:PD_key}(d)), hiding the
underlying secret
key from him/her.

\subsection{Formal security analysis} 
\label{sec:security_analysis}
   We formally define the problem of split manufacturing and highlight the underlying notion of security. Without
loss of generality, we assume $n$ inputs, $m$ outputs, and $k$ key-bits.
\begin{definition}
Let a combinational circuit
be denoted as $C$, where $C$ implements a Boolean function $F: I \xrightarrow{} O$;
$I \in \{0,1\}^n$ and $O \in \{0,1\}^m$, i.e., $C(x,i) = F(i) \forall i\in I$, where $x$ are the combinational elements of $C$.\footnote{We
formalize here for only combinational circuits, but the notion can be readily extended for sequential designs.}
Thus,
a split manufacturing scheme $\mathcal{S}$ can be defined as follows:
\begin{enumerate}
    \item A split procedure is a function $\mathcal{G}: C(x) \xrightarrow{} \{C(x_1, x_2), \lambda(x_2)\}$, where $x_1$ denotes the
    combinational elements whose connections are complete, whereas $x_2$ denotes the elements which are left unconnected. $\lambda(x_2)$
    contains the connectivity information for $x_2$.
    \item $C(x_1, x_2)$ is outsourced to FEOL, whereas $\lambda(x_2)$ is completed at a trusted BEOL, i.e., it remains the secret.
    \item A circuit is compiled by completing $\lambda(x_2)$ connections on $C(x_1, x_2)$ which can be viewed as a function
    $\mathcal{H}:\{C(x_1, x_2), \lambda(x_2)\}\xrightarrow{} C_h(x)$ such that $C_h(x,i) = C(x,i), \forall i\in I$.
\end{enumerate}
\end{definition}

Hence, the security relies on successfully hiding $\lambda(x_2)$ from the {\em untrusted} FEOL foundry.
The success of the attacker can be measured by the difficulty of his/her ability to recover $\lambda(x_2)$ from $C(x_1, x_2)$.
Let us assume an attacker $\mathcal{A}^{\Delta}$, following an attack strategy $\Delta$, tries to recover $\lambda'(x_2)$.
We define the attack success if and only if
$\lambda'(x_2) \equiv \lambda(x_2)$, such that

\vspace{-.8em}
\small
\begin{eqnarray*}
\forall i\in I: && C'_h(x,i) = C(x,i);\\
C'_h(x_1, x_2) = \mathcal{H}(C(x_1, x_2), \lambda'(x_2)): && \lambda'(x_2) \xleftarrow{} \mathcal{A}^{\Delta}(C(x_1, x_2)) 
\end{eqnarray*}

\normalsize
\begin{definition}
A split manufacturing scheme $\mathcal{S}$ is considered to be secure if, for any probabilistic polynomial time (PPT) attacker, the
probability of finding
$\lambda'(x_2) \equiv \lambda(x_2)$ is not greater than $\epsilon(\gamma)$, i.e.,
\begin{equation*}
Pr [\lambda'(x_2) \equiv \lambda(x_2)] \leq \epsilon(\gamma)   
\end{equation*}
where {a function $\epsilon$ is negligible
iff $\forall c\in N, \exists \gamma_0\in N$ such that $\forall \gamma \geq \gamma_0, \epsilon(\gamma) < \gamma^{-c}$, with
$\gamma$ being the security
parameter.}
\end{definition}
The following theorem establishes the security of our proposed scheme against proximity attacks.
\begin{theorem}
Our proposed scheme is secure against a PPT attacker following the strategy of~\cite{wang18_SM}, denoted as $\Psi$, i.e.,
\begin{equation*}
    Pr [\lambda'(x_2) \equiv \lambda(x_2)] \leq \epsilon(\gamma); \quad\lambda'(x_2) \xleftarrow{} \mathcal{A}^{\Psi} (C(x_1, x_2))
\end{equation*}
\end{theorem}

{\em Proof outline.} The success of any proximity attack hinges on FEOL-level hints that can be exploited to infer the missing BEOL connections.
Specifically, the state-of-the-art attack in~\cite{wang18_SM} discusses 1) physical proximity between connected cells, 2) routing of nets in the FEOL, 3)
load constraints for drivers, 4) absence of combinational loops, and 5) timing constraints.\footnote{As for other attacks (e.g., the machine-learning-based attack in~\cite{zhang18}, which was not available to us at the time of writing),
we believe that our security-centric design of the key renders any FEOL-based hint on the key-bits void.
}
It is important to understand that none of the above hints apply to the TIE cells and key-gates in our scheme:
\begin{enumerate}
    \item {\bf Physical proximity} between TIE cells and key-gates are eliminated by randomizing the placement of TIE cells.
    
    \item    {\bf FEOL routing} for the key-nets is completely eliminated by lifting whole key-nets to the BEOL
(Fig~\ref{fig:PD_key}; see also
		    Sec.~\ref{sec:PD_framework}).
    This way, any FEOL wiring which might otherwise leave hints is avoided to begin with.
    
     \item   {\bf Load capacitance constraints} are not applicable to TIE cells, since they are not actual drivers.
    
    \item {\bf Combinational loops} are absent from any key-net path by default, since a TIE cell is not driven by another gate.
    Thus,
    trying to avoid loops for key-nets cannot help an attacker for ruling out incorrect connections.

    \item {\bf Timing constraints} do not apply to TIE cells, which define only static paths for the key-nets. 
    
\end{enumerate}

In the following proof, we leverage the assumptions from the proof outline above.
As a consequence, an attacker is forced
to brute-force the key, which becomes exponentially hard in the number of key-bits used to lock the design.
\begin{proof}
The probability of guessing the correct key-bit for each key-gate is {$P_{kb} \leq \frac{1}{2} + \epsilon$}. Let us assume the probability of finding the correct
connection for all regular nets is $P_o$.
Thus, the probability of successfully recovering all connections is

\vspace{-.8em}
\small
\begin{equation*}\label{eq:sec}
{Pr[\lambda'(x_2) \equiv \lambda(x_2)] = P_o \times P_{kb} \leq P_{kb} = \prod_{i=1}^k \left(\frac{1}{2} + \epsilon\right) \leq \epsilon(k)}
\end{equation*}
\normalsize
where $k$ is the number of key-bits. Note that as the security of our scheme depends only on the key-nets, not the regular nets, we ignore $P_o$.\footnote{
In practice, finding the correct connections for all the regular nets poses a significant
	challenge to the attacker; see also Section~\ref{sec:exp_security}. Nevertheless, following Kerckhoff's principle, we base our formalism only on the key-nets.}
Thus, our scheme is secure for a sufficiently large number of key-bits~\cite{security_key_size}.
This concludes the proof.
\end{proof}

Alternatively, an attacker may want to resort to key extraction attacks commonly leveraged against logic locking, in particular SAT attacks, e.g.,~\cite{subramanyan15}.
However, recall the
\emph{absence of an oracle} for our scheme, and further note that locking has been established as {\em algorithmically secure} in such an
oracle-less model~\cite{epic}.
Thus, such attacks are deemed futile.

\section{Physical Design Framework}
\label{sec:PD_framework}
\begin{figure}[t!]
	\centering
	\includegraphics[width=.5\textwidth, trim = {3.5cm 0 3cm 0}, clip = true] {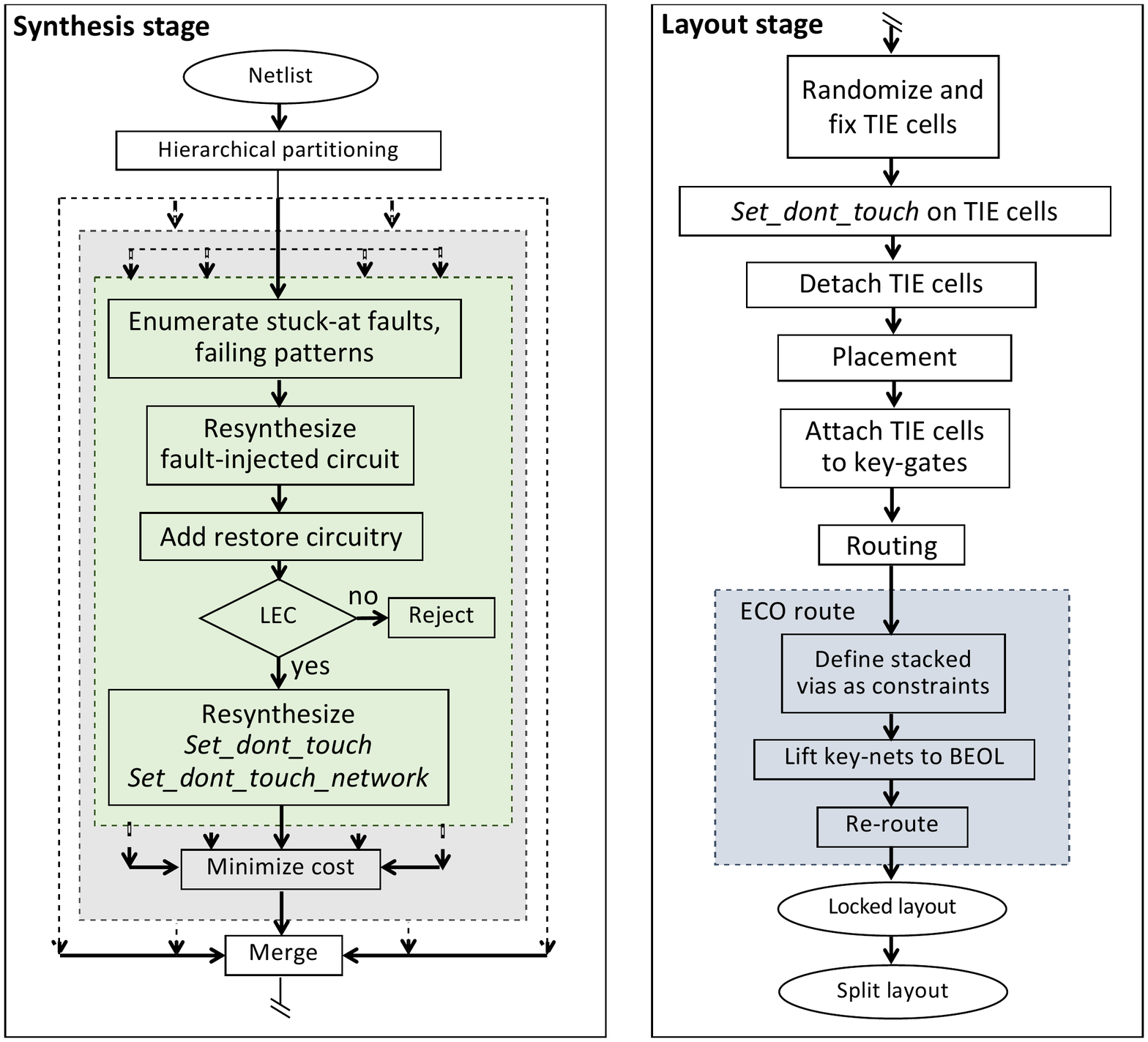}
	\caption{
		Our physical design flow,
		implemented using \emph{Synopsys Design Compiler}, \emph{Cadence Innovus}, and custom code.
		For the synthesis stage, dashed arrows indicate parallel processing. Furthermore, \emph{Cadence LEC} is leveraged
			to confirm the equivalence with the original netlist during locking.
	}
	\label{fig:flow}
\end{figure}

We would like to emphasize the fact that our scheme is generic and agnostic to the underlying locking technique.
For a meaningful case study, however,
    we extend a recent locking technique~\cite{sengupta_vts18}.
Figure~\ref{fig:flow} illustrates the flow; details are given next.
We also provide our flow to the community~\cite{webinterface}.

\subsection{Security-centric and cost-driven synthesis stage}
Besides locking,
the other crucial concept
in our scheme---lifting of key-nets
to the BEOL---may incur considerable cost, as shown, e.g., in~\cite{patnaik_aspdac18}.
To limit such potential cost, we adapt and extend the technique presented by Sengupta \emph{et al.}~\cite{sengupta_vts18}.
Doing so even allows us to obtain \emph{area savings} when compared to the original, unprotected
designs (Sec.~\ref{sec:exp_layout}).
The key steps in~\cite{sengupta_vts18} are to (1)~selectively introduce some stuck-at faults in the original circuit, (2)~re-synthesize the
circuit to remove the stuck-at logic parts, (3)~insert some restore circuitry to enable recovery from those faults,
and (4)~configure the restore circuitry with a secret key, by making key-gates an essential part of the restore circuitry.

\begin{figure}[t!]
	\centering
	\includegraphics[width=.97\columnwidth] {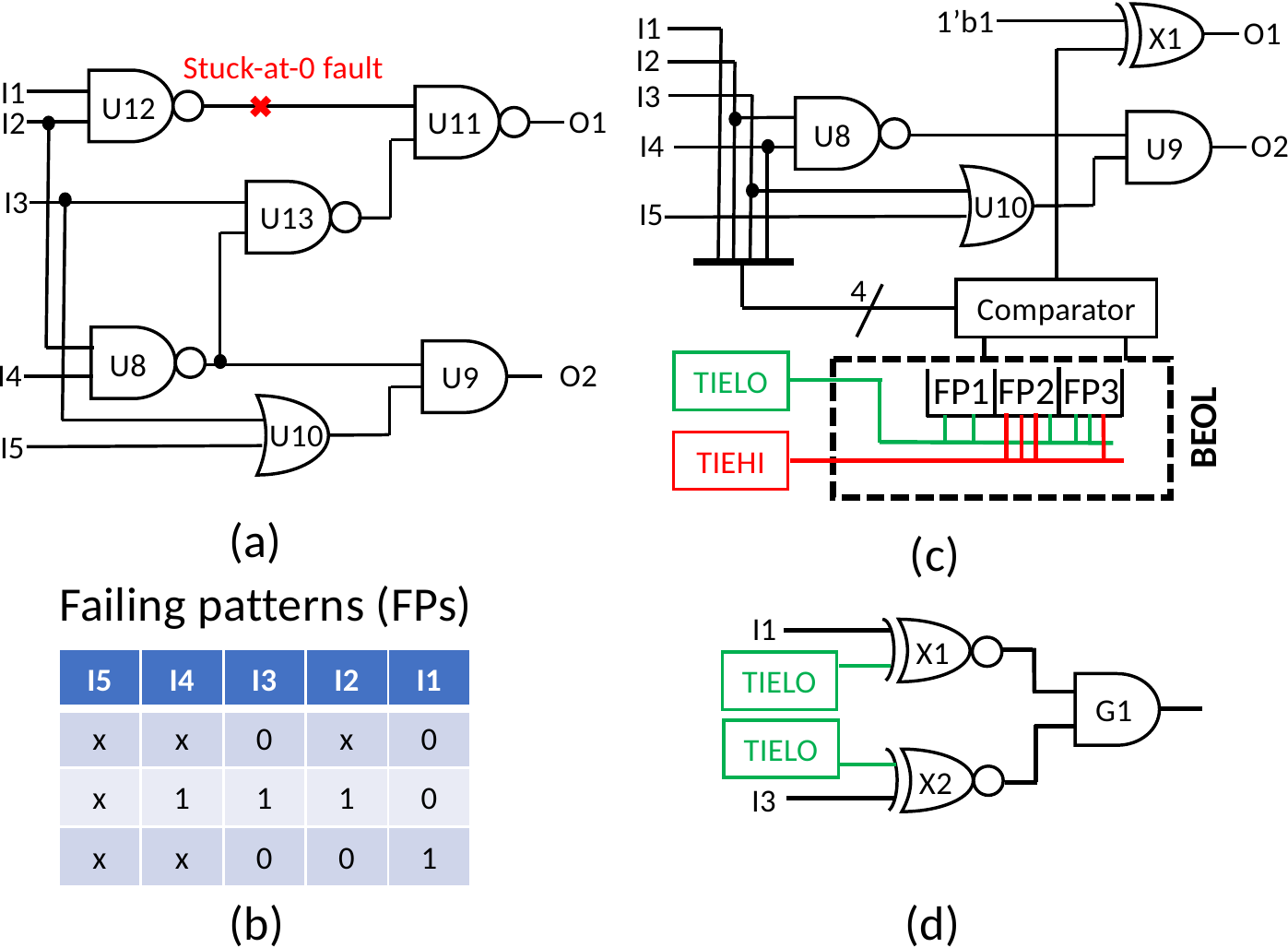}  
	\caption{
		Simple example on \emph{c17} for the fault-injection based 
        technique used in our scheme.
		(a) The original circuit, with a fault injected at the output of U12.
		(b) List of related failing patterns.
		(c) Locked circuit with the key-nets lifted to the BEOL.
		(d) Comparator for the first failing pattern.
	}
	\label{fig:exmpl}
\end{figure}

We largely re-implement these steps, but also extend and tailor them specifically
for our scheme.
Next, we provide some details; an illustrative example is given in Fig.~\ref{fig:exmpl}.

Stuck-at faults and their failing patterns can be fully enumerated using automatic test pattern generation (ATPG) tools,
but doing so can impose considerable computational cost, depending on the size of the design to protect.
Therefore, we initially partition the netlist
in a random but balanced manner.
Then, we can explore the stuck-at faults for all
modules independently.
Besides \emph{parallel processing}, this allows us to protect all the parts of the design.
For each of the
modules, we pick its most cost-effective failing pattern; see below for the cost model.
The selected patterns are then provided to the restore circuitry, which is essentially a comparator along with the key-gates.
Hence, the key-nets can also be thought of as a hard-coded input for the restore circuitry.
To avoid re-structuring of the key-nets by the design tools,
we apply the specific commands \emph{set\_dont\_touch} and {\em set\_dont\_touch\_network} on the TIE cells and key-nets, respectively.

The number of key-bits/failing patterns dictates the layout cost by the following trade-off:
the more faults are injected, the more logic can be
removed, but the larger becomes the restore circuitry.
We consider the total cell area, including the restore circuitry, to evaluate cost during the synthesis runs:\footnote{
The cost model
can be readily extended to power and/or timing, if desired.}

\vspace{-.8em}
\footnotesize
\begin{equation*}
    cost = \min_{f\in F} \{cost_{f}^{fi} + cost_{f}^{rest} \} \mbox{ such that }
    \begin{cases}
    		    |\mathbf{K}| = k\\
		    \mathbf{K} \xleftarrow{\$} \{0,1\}^k
	\end{cases}
	\vspace{-1ex}
\end{equation*}
\normalsize
where $F$ is the list of faults, $cost_f^{fi}$ denotes area cost for the fault-injected circuit, $cost_f^{rest}$ denotes area cost for
the restore circuitry, and $\mathbf{K}$ denotes the key. The constraint $|\mathbf{K}| = k$ ensures that the restore circuitry contains $k$ key-bits, thereby, ensuring adequate and controllable security.
The constraint $\mathbf{K} \xleftarrow{\$}
\{0,1\}^k$ ensures
a uniform bit distribution for the
key $\mathbf{K}$,
which directly translates to an even distribution of TIEHI and TIELO cells in the FEOL.
Thus, an attacker cannot derive hints from the
distribution of TIE cells.

\subsection{Secure layout generation}\label{sec:layout}
As indicated, randomizing the placement of TIE cells and lifting the key-nets to the BEOL are essential steps in our
scheme. Toward this end, we randomize and fix the placement of TIE cells and use {\em set\_dont\_touch} on the TIE cells. Before regular
placement of the locked design, TIE cells are detached from the key-gates to avoid inducing any layout-level hints. Afterwards, key-gates
are re-attached before routing, and we apply ECO routing to lift the key-nets to the BEOL.
For that, we enforce routing of key-nets as new nets, while other regular nets are re-routed if needed.
We declare routing constraints such that \emph{stacked vias} are used directly at the output pins of all TIE cells and the input pins
of all key-gates. These constraints ensure that whole key-nets are lifted to the BEOL at once, and they can be configured for different
split layers.

\section{Experimental Investigation}\label{sec:results}
In this work, we use 128 key-bits, which is deemed secure according to modern security standards~\cite{security_key_size}.
All experiments have been carried out on an 128-core Intel Xeon processor running at 2.2 GHz with 256 GB of RAM.
We use Synopsys Design Compiler (DC) and Cadence Innovus along with the 45nm
Nangate OpenCell library~\cite{nangate11} for layout generation.
For fault simulations, we use the Atalanta-M ATPG tool, which is able to provide all failing patterns.
Also, we use Cadence Conformal LEC to verify the equivalence of the original design and the locked design. That is required since
Atalanta does not guarantee formal equivalence by itself.

Runtimes for our experiments on the large-scale ITC'99 benchmarks (containing tens of thousands of gates) are in the range of 5--18
hours. The most time-consuming aspect for our flow are the re-synthesis runs using DC.
Given the parallel architecture of our flow, these runtimes could be significantly reduced once a sufficient number of DC licenses
is available.

\subsection{Security analysis}
\label{sec:exp_security}
Besides the formal analysis in Sec.~\ref{sec:security_analysis}, we use the attack of~\cite{wang18_SM}
for an empirical study.
The attack as is may falsely connect a key-gate 
to a regular driver.
Since an attacker can understand which gates are key-gates from the FEOL,
we
customize/improve the attack as follows.
For any key-gate being falsely connected to a regular driver, we re-connect this key-gate to a TIEHI or TIELO cell in a random manner (but key-gates already connected to a TIE cell are kept as is).

\noindent\textbf{Correct connection rate (CCR).}
CCR measures
the ratio of correctly inferred connections to that of the total number of
broken connections;
the lower the CCR, the better the protection.
Since the security of our scheme depends on the key-nets, we report CCR for regular nets and key-nets separately.
Further, we differentiate between physical and logical CCR for key-nets:
physical CCR concerns 
whether the original routing from the particular TIE cell to the particular key-gate is correct, logical CCR concerns whether
a particular key-gate is connected to any TIE cell of correct logical value.
From a designer's point of view, 
the logical CCR for key-nets should be $\sim$50\%, implying the attacker cannot do better than random guessing (Sec~\ref{sec:security_analysis}).

\begin{table}[tb!]
\scriptsize
\setlength{\tabcolsep}{0.33em}
\centering
\caption{CCR (\%) for ITC'99 benchmarks when split at M4 and M6, ``NA'' means attack time-out after 72h}
\label{tab:ccr}
\begin{tabular}{|c|c|c|c|c|c|c|} 
\hline
 \multirow{3}{*}{\bf Benchmark} &  \multicolumn{3}{c|}{{\bf M4}} &  \multicolumn{3}{c|}{{\bf M6}} \\ \cline{2-7}
 &  \multicolumn{2}{c|}{{\bf Key nets}} & \multirow{2}{*}{\bf Regular nets} & \multicolumn{2}{c|}{{\bf Key nets}} & \multirow{2}{*}{\bf Regular nets} \\ \cline{2-3} \cline{5-6}
 & {\bf Logical} & {\bf Physical} &  & {\bf Logical} & {\bf Physical} & \\ \hline \hline		
b14	& 52 & 1 & 17 & 54 & 2 & 47 \\   \hline
b15	& 49 & 0 & 15 & 49 & 0 & 25 \\    \hline
b17	& NA & NA& NA & 51 & 1 & 21 \\	\hline
b20	& 54 & 0 & 17 & 60 & 0 & 36 \\	\hline
b21	& 50 & 0 & 14 & 54 & 0 & 36 \\	\hline
b22	& 52 & 0 & 14 & 55 & 0 & 25 \\	\hline
{\bf Average}	& {\bf 51} & {\bf 0} & {15} & {\bf 54} & {\bf 1} & {\bf 32}\\	
\hline
\end{tabular}
\end{table}

In Table~\ref{tab:ccr}, we report CCR for two different setups, namely for lifting of key-nets
to M5 and M7 while splitting at M4 and M6, respectively.
As can be expected for regular nets~\cite{wang18_SM}, CCR improves for higher split layers.
The physical CCR for key-nets, however, is close to zero for all cases.
This confirms our claim of a physically secure key design.
Moreover, the logical CCR is $\sim$50\%, i.e., the attack cannot perform better than any random guess.\footnote{Recall that we 
post-process falsely connected key-gates from~\cite{wang18_SM}. Otherwise, as we find in separate experiments, the logical CCR
drops well below 50\%, namely to 29.3\% and 17.6\% for split layers M6 and M4, respectively.}
Also note that \emph{any deviation from 50\% cannot be leveraged by an attacker}: he/she cannot know which
particular key-bits are correct/wrong without using an oracle, and is thus left with a number of possible choices that is exponential in the number of key-bits.
Finally, another important observation is that the logical CCR is similar for both split layers. This establishes the fact that the
security of our scheme is {\em agnostic to the split layer}, i.e., \emph{key-nets can be split at any layer without providing any
further benefit than random guessing} does for the attacker.

\noindent\textbf{Hamming distance (HD) and output error rate (OER).}
HD quantifies the
difference for the output between the original netlist and the one recovered by the attacker.
From the defender's perspective, the ideal HD is $\sim$50\%.
OER measures the likelihood of any output error in the netlist recovered by the attacker;
the higher the OER, the better the protection.

From Table~\ref{tab:hd_oer}, we see that the proximity attack is unable to recover the functionality of the original netlist.
While HD is close to 50\% for the layouts split at M4, we note that HD drops for the layouts split at M6.
That is because when splitting at a higher layer, an attacker can readily obtain a larger part of the design from the FEOL via the regular nets.
For the overall
scheme, it is important to recall that the OER is 100\% and logical CCR for key-nets is $\sim$50\% even for higher layers, so the designs remain protected.
Independently, the designer may increase the number of key-bits to raise the HD.

To validate that our scheme remains secure even in the presence of an ``ideal proximity attack,'' we also conduct the following experiment.
The baseline here is that we assume \emph{all} regular nets have been correctly inferred; only key-nets remain to be attacked.
As demonstrated above and throughout vast prior art, this represents a strong assumption to begin with.
Now, as established in Sec.~\ref{sec:security_analysis} and empirically confirmed above, an attacker cannot do better than randomly
guessing the key-nets. Therefore, we apply 1,000,000 runs for randomly guessing the key-nets.
For these experiments, the OER remains at 100\% across all benchmarks, \emph{establishing the security of our technique even in the presence
	of such an ideal attack}.

\begin{table}[tb!]
\scriptsize
\centering
\caption{HD and OER (\%) for ITC'99 benchmarks when split at M4/M6, for 1M simulation runs, ``NA'' means attack time-out after 72h}
\begin{tabular}{|c|c|c|c|c|} 
\hline
 \multirow{2}{*}{\bf Benchmark} &  \multicolumn{2}{c|}{{\bf M4}} &  \multicolumn{2}{c|}{{\bf M6}}
\\ \cline{2-5} & {\bf HD} & {\bf OER} & {\bf HD} & {\bf OER} 
 \\ \hline\hline
b14	& 46 & 100 & 25 & 100 \\ \hline
b15	& 52 & 100 & 20 & 100 \\  \hline
b17	& NA & NA  & 31 & 100 \\  \hline
b20	& 57 & 100 & 19 & 100 \\	\hline
b21	& 56 & 100 & 26 & 100 \\	\hline
b22	& 57 & 100 & 27 & 100 \\	\hline
{\bf Average}	& {\bf 53} & {\bf 100} & {\bf 25} & {\bf 100}\\	
\hline
\end{tabular}
\label{tab:hd_oer}
\end{table}

\noindent\textbf{Comparison with prior art.}
Recall that our technique, unlike prior art, offers formal  security guarantees for the first time concerning the classical threat model. Further, we validate these formal claims empirically by running~\cite{wang18_SM}.
Nevertheless, for a meaningful comparative study, we assess our work against recent prior art (Tab.~\ref{tab:comp}). As preset by the prior art, here we leverage the ISCAS benchmarks and present CCR, HD, OER, and percentage of netlist recovery (PNR).
PNR measures the structural similarity between the protected netlist and the one obtained by the attacker~\cite{patnaik_aspdac18}; the lower the PNR, the better the protection.
Note that CCR for ours refers to the physical CCR of the key-nets. It is evident from Table~\ref{tab:comp} that ours is competitive or even superior to all the latest schemes.

\begin{table*}[tb!]
\vspace{-0.7em}
\scriptsize
\centering
\caption{PNR, CCR, HD, and OER (all in \%) for ISCAS benchmarks when split at M4, ``NA'' means not reported in the respective publication}
\label{tab:comp}
\begin{tabular}{|c|c|c|c|c|c|c|c|c|c|c|c|c|c|c|c|c|} 
\hline
 \multirow{2}{*}{\bf Benchmark} &  \multicolumn{4}{c|}{{\cite{wang17}}} &
 \multicolumn{4}{c|}{{\cite{patnaik_aspdac18}}} &
 \multicolumn{4}{c|}{{\cite{patnaik_dac18}}}&
 \multicolumn{4}{c|}{{\bf Proposed}}
\\ \cline{2-17}
 & {\bf PNR} & {\bf CCR} & {\bf HD} & {\bf OER} & {\bf PNR} & {\bf CCR} & {\bf HD} & {\bf OER}
 & {\bf PNR} & {\bf CCR} & {\bf HD} & {\bf OER}
 & {\bf PNR} & {\bf CCR} & {\bf HD} & {\bf OER}
 \\ \hline\hline
c432	& 87.5 & 78.8 & 46.1 & 99.4 & 32.3 & 0 & 45. 9& 100 &NA& 0 & 48.4 & 99.9 &28&2&42.5& 98.3\\ \hline
c880	& 86.8 & 45.8 & 18.0 & 99.9 & 28.3 & 0 & 39.9 & 100 &NA& 0 & 43.4 & 99.9 &29&1&35.7&100\\  \hline
c1355	& 84.9 & 77.1 & 26.6 & 100 & 32.8 & 0 & 46.1 & 100 &NA& 0 & 40.1 & 99.9 &31&0&32.3&100\\  \hline
c1908	& 91.2 & 83.8 & 38.8 & 100 & 29.5 & 0 & 48.1 & 100 &NA& 0 & 46.2 & 99.9 &26&1&34.4&100\\	\hline
c3540	& 86.2 & 77.0 & 36.1 & 100 & 30.8 & 0 & 46.4 & 100 &NA& 0 & 47.9 & 99.9 &16&2&37.8&100\\	\hline
c5315   & 87.7 & 74.7 & 18.1 & 100 & 31.6 & 0 & 35.4 & 100 &NA& 0 & 38.3 & 99.9 &31&1&45.2&100\\	\hline
c7552   & 93.9 & 73.9 & 20.3 & 100 & 26.9 & 0 & 25.7 & 100 &NA& 0 & 27.8 & 99.9 &31&1&71.7&100\\    \hline
{\bf Average}	& {\bf 88.3} & {\bf 73.3} & {\bf 29.1} & {\bf 99.9} & {\bf 30.3} & {\bf 0} & {\bf 41.1} & {\bf 100} &{\bf NA}& {\bf 0} & {\bf 41.7} & {\bf 99.9} &{\bf 27.5}&{\bf 1.1}&{\bf 42.8}&{\bf 99.8}\\	
\hline
\end{tabular}
\vspace{-0.5em}
\end{table*}

\subsection{Layout analysis}
\label{sec:exp_layout}

\begin{figure}[t!]
	\centering
	\includegraphics[width=0.33\columnwidth, angle=-90] {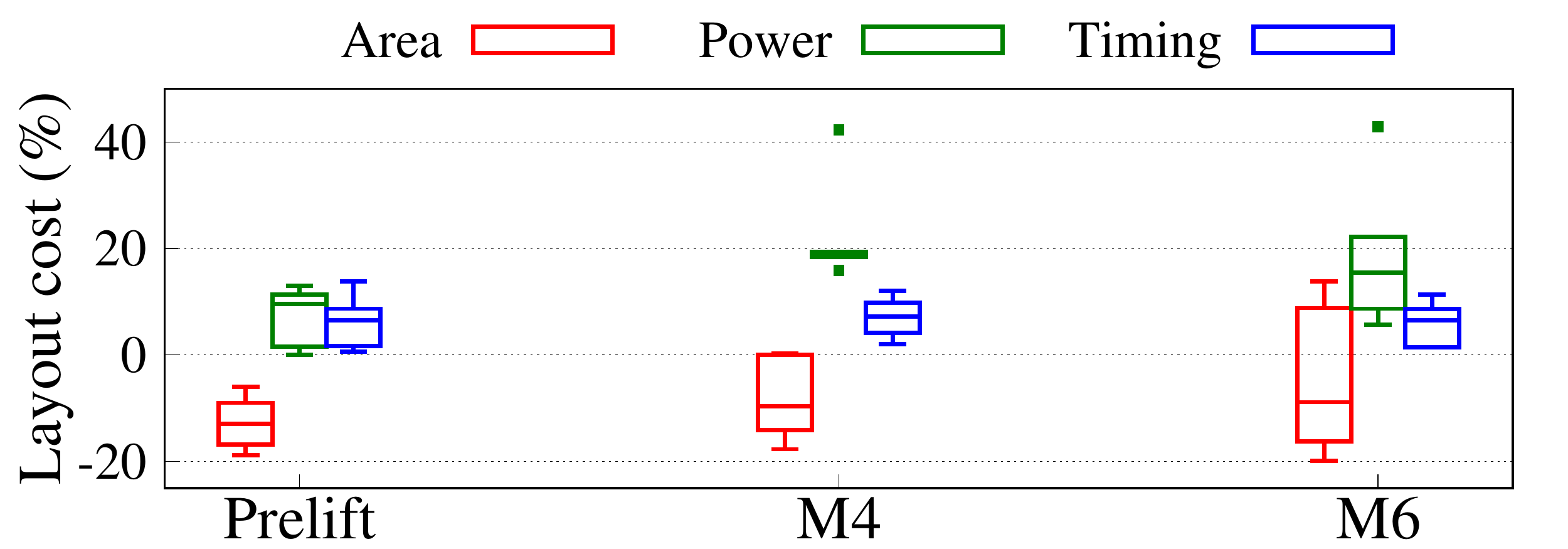} 
	\vspace{5pt}
	\caption{Layout cost for our scheme. The baseline are the unprotected layouts.
	\emph{Prelift} refers to
locked layouts without lifting of key-nets. Each box comprises data points within the first and third quartile, the bar represents the
	median, the whiskers the minimum/maximum values, and outliers are marked by dots.
	}
	\label{fig:ppa}
\end{figure}

Figure~\ref{fig:ppa} illustrates the layout cost across all considered ITC'99 benchmarks.  The general baseline are the
regular, unprotected layouts.
We ensure that all regular and locked layouts have only few, if any, outstanding DRC issues that can be manually fixed.
Toward that end, we reduce the utilization rates as needed.
Hence, area is reported in terms of die outline.

\emph{Prelift} serves as a crucial reference point:
it covers the locked layouts as generated using a regular physical design flow, but with TIE cells and key-nets marked as ``don't
touch'' (i.e., Fig.~\ref{fig:PD_key}(a)).
We note considerable \emph{area savings} in contrast with the regular layouts,
namely 12.75\% on average.
This is attributed to the underlying principle of removing logic
being tailored for area saving in this work (Sec~\ref{sec:PD_framework}).
We would like to re-emphasize
	that all locked layouts are functionally equivalent to the original netlist.
Naturally, such area savings enforce some trade-off: power and timing are increased by 7.66\% and 6.40\% on average,
respectively.

Next, we discuss the cost for the final layouts when compared with the unprotected layouts.
Notably, \emph{area savings} still carry over, namely with 8.83\% and 10.05\% for splitting at M6 and M4, respectively.
Power is increased by 15.46\% and
20.34\% when
splitting at M6 and M4, respectively.
That is due to the fact that lifting of key-nets (using stacked vias) enforces some re-routing of regular nets. This, in turn, requires
upscaling of drivers and/or insertion of buffers to meet timing (applies only to regular nets, not key-nets).
For timing, cost is thus limited to 6.53\% and 6.25\% for M6 and M4,
respectively.
In short, our secure PD flow for split manufacturing imposes acceptable power and timing cost along with area savings.

\noindent\textbf{Comparison with prior art.}
Few works report on layout cost incurred by their protection scheme.  Recently, \cite{patnaik_aspdac18,patnaik_dac18} report layout cost, but for ISCAS
benchmarks which contain only few hundreds of cells. On average, these schemes incur
9.2\%/10.7\%/15.0\%~\cite{patnaik_aspdac18} and 0.0\%/11.5\%/10.0\%~\cite{patnaik_dac18} cost for area/power/delay, respectively.
The
cost for our scheme is competitive with this prior art in terms of delay, and we can obtain even area savings which
this prior work cannot, all the while securing practically more relevant
ITC'99 benchmarks.\footnote{We may incur higher
cost for ISCAS benchmarks (e.g.,
34.8\%/70\%/1.6\% for c7552 for area/power/timing). Due to the small size of those designs, the key-gates and restore circuitry form a considerable part.
However, cost are amortized for larger designs, as shown for ITC'99 benchmarks.  We can also argue that protecting overly small designs is not meaningful to begin with.}

\section{Conclusion and Future Work}

For the first time, we present a formally secure scheme for split manufacturing concerning the classical threat model.
Our paradigm is to lock the FEOL by embedding a secret key; this is in fundamental contrast to current defense schemes which all
rely on heuristic protection techniques (i.e., layout-level perturbations). 
The secret key to unlock the design needs to be implemented at the BEOL. We develop and openly release a design flow to embed the key such
that it becomes indecipherable to an FEOL-centric proximity attack.
At the same time, our flow can provide \emph{cost savings} in terms of reduced die outlines.
While we cannot foresee future proximity attacks, e.g., based on advanced machine learning, we believe our scheme will
remain resilient. Any proximity attack has to rely on FEOL-level hints, and such hints are inherently avoided for
the secret key by our core techniques of randomizing TIE cells and lifting the key-nets in full.

We present extensive results on large-scale ITC'99 benchmarks that further validate our formal claims.
Two notable findings are as follows.
First, we show that our scheme is secure against a state-of-the-art proximity attack, which cannot perform better than randomly guessing the
key bits.
Second, the resilience of key-nets is independent of the split layer.

For that latter finding, we propose---for future work---a scenario where a trusted packaging facility replaces the trusted BEOL fab.
As the security of our approach stems from hiding the bit assignments for the key-nets, these nets can also be connected to the IO ports of a chip and, in turn, tied to
fixed logic at the (trusted) package routing level.

\section*{Acknowledgments}
This work is partially sponsored by the New York University/New York University Abu Dhabi (NYU/NYUAD) Center for Cyber Security (CCS).
We also thank Dr.\ Anja Henning-Knechtel for preparing Fig.~\ref{fig:flow}.

\tiny
\setstretch{0.8}

\end{document}